\documentclass[aps,pra,superscriptaddress,twocolumn,longbibliography,amsmath,amssymb]{revtex4-1}

\usepackage{graphicx}
\usepackage{braket, color, physics}
\usepackage[colorlinks = true, urlcolor = black, citecolor = blue, linkcolor = red]{hyperref}
\newcommand{\half}{\frac{1}{2}}

\begin{document}

\title{Noiseless attack and counterfactual security of quantum key distribution}
\author{Vinod N. Rao}
\email{vinod@ppisr.res.in}
\affiliation{Theoretical Sciences Division, Poornaprajna Institute of Scientific Research, Bengaluru - 562164, India}
\affiliation{Graduate Studies, Manipal Academy of Higher Education, Manipal - 576104, India}
\author{R. Srikanth}
\email{srik@ppisr.res.in}
\affiliation{Theoretical Sciences Division, Poornaprajna Institute of Scientific Research, Bengaluru - 562164, India}
 
\begin{abstract}
Counterfactual quantum key distribution (QKD) enables two parties to share a secret key using an interaction-free measurement. Here, we point out that the efficiency of counterfactual QKD protocols can be enhanced by including non-counterfactual bits. This inclusion potentially gives rise to the possibility of noiseless attacks, in which Eve can gain knowledge of the key bits without introducing any errors in the quantum channel. We show how this problem can be resolved in a simple way that naturally leads to the idea of ``counterfactual security'', whereby the non-counterfactual key bits are indicated to be secure by counterfactual detections. This method of enhancing the key rate is shown to be applicable to various existing quantum counterfactual key distribution protocols, increasing their efficiency without weakening their security.
\end{abstract}

\maketitle

\section{Introduction \label{sec:intro}}

Quantum key distribution (QKD) promises the security of secret communication based only on quantum mechanical no-go theorems such as no-cloning and the impossibility to perfectly distinguish non-orthogonal states, rather than computational assumptions, as is the case with conventional public key crypto-systems. Since the early schemes \cite{BB84,ekert1991quantum, goldenberg1995quantum}, several QKD protocols have been proposed. Counterfactual QKD protocols \cite{guo1999quantum, noh2009counterfactual, sun2010counterfactual, shenoy2013semi, salih2013protocol, vaidman2019analysis} are based on the principle of interaction-free measurement (IFM) \cite{elitzur1993quantum}, whereby a key bit is generated even without the physical transmission of the particle, in the sense that the block actions by one party (Bob) are revealed by detections of another party (Alice).

The principle of IFM has been employed for other quantum information processing tasks beside cryptography, such as quantum computation \cite{cao2020counterfactual, li2020counterfactual}, entanglement generation \cite{shenoy2015counterfactual}, direct communication \cite{salih2013protocol,aharonov2019modification} and the device-independent (DI) version of counterfactual QKD \cite{kamaruddin2020counterfactual} as well. For a recent review of QKD as well as other aspects of quantum cryptography, see Ref. \cite{shenoy2017quantum}.

Typically in a QKD scheme, a fraction of potential key bits (detection data that would lead to key generation) is sacrificed as check bits in order to estimate the the quantum bit error rate (QBER) $e$. Analogously in a counterfactual QKD scheme, a fraction of the secret bits shared via IFM (the so-called counterfactual bits) are used up as check bits. Here, we introduce a twist to this paradigm. We propose to use the non-counterfactual bits as key bits, the security of which is proven by counterfactual statistics. In such a case, these non-counterfactual key bits are said to be ``counterfactually secure''.

This idea is based on the observation that usually counterfactual QKD protocols \cite{noh2009counterfactual, guo1999quantum, shenoy2013semi} discard non-counterfactual bits. Our motivation is to improve the efficiency of these protocols, by including the non-counterfactual bits into the key. However, a direct inclusion of these bits is shown to lead to the possibility of a noiseless attack by the eavesdropper Eve, whereby she gains partial or full knowledge of key bits without introducing errors that Alice and Bob can detect in the quantum channel. To counter this, it suffices for Alice and Bob to perform probabilistic spin flip operations, which-- interestingly-- lead to counterfactual security.

Counterfactuality has been applied in other areas of quantum information processing, such as entanglement generation \cite{shenoy2015counterfactual}, quantum computation \cite{cao2020counterfactual, li2020counterfactual} and to direct communication \cite{salih2013protocol, gisin2013optical, aharonov2019modification}, where the definition of counterfactuality, based on a weaker or more stringent criterion, has been debated \cite{vaidman2014comment, salih2014salih, vaidman2019analysis, hance2019quantum}. Counterfactual QKD has been experimentally implemented \cite{brida2012experimental, ren2011experimental} as well.

The present work is organized as follows. The prototypical counterfactual quantum QKD protocol \cite{noh2009counterfactual}, namely Noh09, is briefly presented in Sec. \ref{sec:proto}. The inclusion of the non-counterfactual key bits in this protocol is proposed as a way to improve efficiency. We point out that this opens up a new type of eavesdropping, the so-called noiseless attack. A further modification of the protocol to thwart this attack is then given. Sec. \ref{sec:secu} presents the security of the modified protocol, highlighting a novel, counterfactual aspect of it. The analogous extension of other counterfactual QKD protocols by the inclusion of non-counterfactual bits for key is discussed in Sec. \ref{sec:other}. Finally, we conclude in Sec. \ref{sec:conc}.

\section{Non-counterfactual bits and noiseless attacks\label{sec:proto}} 

\begin{figure}[h]
\includegraphics[scale=0.375]{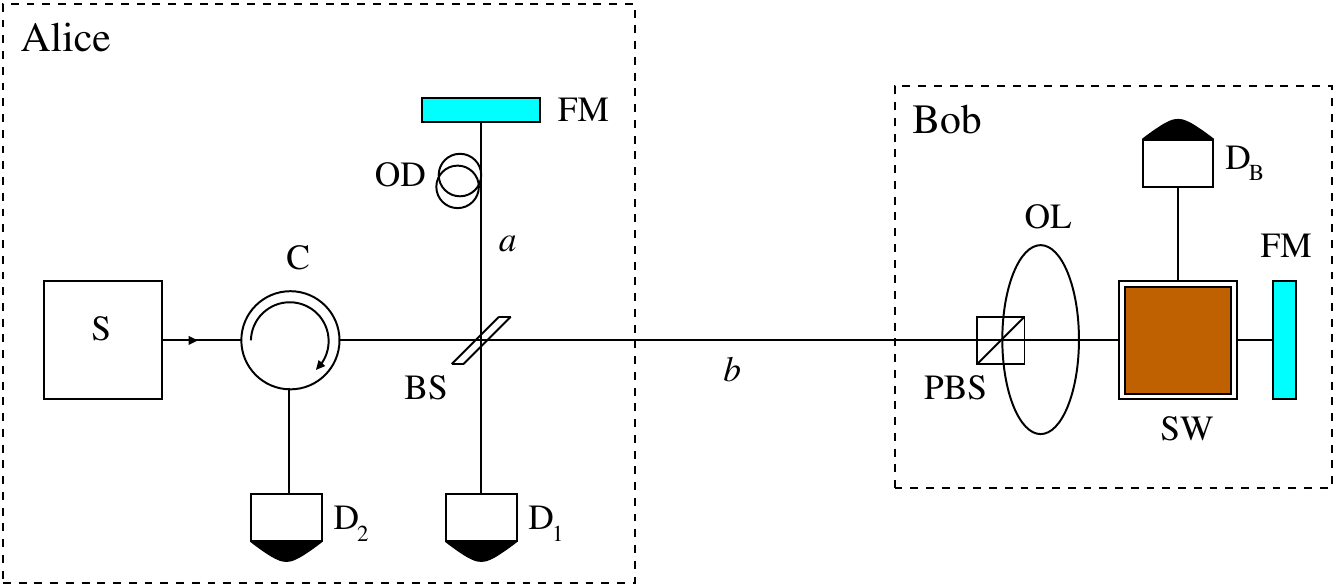}
\caption{(Color online) Experimental setup: Alice injects single photons prepared in polarizations $H$ or $V$ into a Michelson interferometer, on one of whose arm ends Bob is located. He applies a polarization-dependent ``reflect'' or ``absorb'' operation on each photon. Alice and Bob independently flip the polarization of a fraction $f$ of total reflected particles. A detection at $D_1$ represents either an IFM or non-interference due to only one of particles from the two arms being polarization flipped, whereas one at $D_B$ signifies that the photon physically traversed the channel. The detected polarization corresponds to the secret bit.}
\label{fig:Noh09}
\end{figure}

The Noh09 protocol works as follows: (1) Alice transmits single photons prepared in the polarization state $H$ (horizontal) or $V$ (vertical) to a beam-splitter (BS) of a Michelson interferometer M. One of the output arms of M remains internal in her laboratory, whilst the other (external) arm stretches out to the station of Bob (Fig. \ref{fig:Noh09}). The state of the photon after BS is:
\begin{equation}
\ket{\Psi}_{ab} = \frac{1}{\sqrt{2}}(\ket{0,j}_{ab}+ \ket{j,0}_{ab}),
\label{eq:noh}
\end{equation}
where $j \in \{H, V\}$ and $\ket0$ denotes the vacuum state. (2) In each round, Bob may either reflect the photon of polarization $H$ whilst blocking one with polarization $V$ (action $R_H$), or vice versa (action $R_V$), with equal probability. (3) If Bob's action $R_k$ ($k \in \{H,V\}$) matches with the polarization $j$ of the photon (i.e., $j=k$), then it is reflected and detected deterministically at Alice's detector $D_2$. (4) In case of a mismatch (i.e., $j \ne k$), then there are three possibilities of its detection: (i) at $D_B$ in Bob's station (with probability $\half$); (ii) at $D_2$ (with probability $\frac{1}{4}$); (iii) at $D_1$ in Alice's station (with probability $\frac{1}{4}$). (5) Alice publicly announces the $D_1$ detection instances, which constitutes the sifting process. The polarization of the corresponding photon forms the (counterfactual) secret bit. (6) On a fraction of the sifted key, they announce their respective actions to estimate the QBER $e$. If $e$ is too large, they abort the protocol. 

The QBER in the channel is estimated by
\begin{equation}
e \equiv P(H, R_H|D_1) + P(V, R_V|D_1),
\label{eq:noherror}
\end{equation}
where $P(\cdot|\cdot)$ denotes the conditional probability.
Although the encoding states in Eq. (\ref{eq:noh}) are orthogonal, the corresponding states $\rho_j$ accessible to Eve are not. In particular, $\rho_j = \half\left(\ket0_B\bra0 + \ket{j}_B\bra{j}\right),$ whereby the trace distance $D = |\rho_H - \rho_V| = \frac{1}{2}$. Therefore, Eve's optimal probability \cite{holevo1978asymptotically} to guess the correct polarization is $p_{\rm guess} = \half(1+D) = \frac{3}{4}$, so that 
\begin{equation}
e^\prime \equiv 1-p_{\rm guess}=\frac{1}{4}
\label{eq:optEve}
\end{equation}
is Eve's minimum error in distinguishing $\rho_H$ and $\rho_V$.

Now we consider a modified protocol $\mathcal{P}$ which uses $D_B$ detections also for key generation. This is the same Noh09, except that step (5) is replaced by:

($5^\prime$) If no public announcement of counterfactual detection (at $D_1$) is made by Alice, then there should have been a $D_2$ or $D_B$ detection, assuming that the channel is non-lossy, and both Alice and Bob know which happened. (To guard against channel loss, Alice and Bob may publicly discuss to check that their $D_2$ and $D_B$ detections are perfectly anti-correlated on a fraction of these instances.) Only the $D_B$ detections are retained, which constitutes the sifting process. The polarization of the corresponding photon forms the (non-counterfactual) key bit. (Notice that the QBER estimation is done using the counterfactual detections.)

The advantage of this modification is that the efficiency of the protocol $\mathcal{P}$ is doubled (assuming BS to be unbiased), when compared to Noh09,
\begin{equation}
\eta = P_{D_B} = \frac{1}{4},
\label{eq:eta}
\end{equation}
However, as pointed out below, Eve can launch noiseless attack on the protocol, whereby she gets information of all $D_B$ bits without introducing any QBER. 

The state of the particle after the BS operation is given in Eq. (\ref{eq:noh}). Suppose Eve entangles her probe with the attack $\mathcal{U}$, initially prepared in the state $\ket{\varepsilon_0}$, with the particle in the external arm as,
\begin{subequations}
	\begin{align}
		\ket{\alpha, \varepsilon_0}_{be} &\longrightarrow \ket{\alpha,\varepsilon_\alpha}_{be} ~~~ (\alpha \in \{0, H, V\}), \label{eq:onwarda} \\
		\ket{0,\varepsilon_j}_{be} &\longrightarrow \ket{0,\varepsilon_j}_{be} ~~~(j \in \{H, V\}). \label{eq:onwardb}
	\end{align}
	\label{eq:onward}
\end{subequations}
The state of the particle after Eve's attack becomes
\begin{equation}
\ket{\Psi^\prime}_{abe} = \frac{1}{\sqrt{2}}(\ket{j,0,\varepsilon_0} + \ket{0,j, \varepsilon_j})_{abe}.
\label{eq:nohstate}
\end{equation}
and she applies the ``unattack'' $\mathcal{U}^{-1}$ on the return photon mode.

Clearly, Bob cannot test for the coherence of BS arms from $D_B$ detections. If he detects a photon of polarization $j$ by a blocking action, then Eve's probe state is collapsed to $\ket{\varepsilon_j}\bra{\varepsilon_j}$. If Bob's blocking action did not detect a photon, or if he reflected, then Eve's probe is left in the state $\ket{\varepsilon_0}$. Thus Eve can potentially get information on all the $D_B$ bits.

Alice and Bob test for QBER $e$ from counterfactual statistics and Eve can remove her footprint for all Bob-reflected instances, by virtue of her $\mathcal{U}^{-1}$ operation. Hence this attack strategy would be noiseless (i.e., introduces no error). This makes the protocol $\mathcal{P}$ fully insecure. We note that there is no contradiction with the requirement Eq. (\ref{eq:optEve}), since with the key rate in protocol $\mathcal{P}$ is precisely $\frac{1}{4}$. In other words, because of not involving the coherence of the particle, the non-counterfactual bits are not taking advantage of the non-orthogonality of the exposed encoding states.

At first, it seems that including the $D_B$ detections is not a good proposition. Fortunately, there is a simple fix to this problem, while still keeping the efficiency advantage. This will essentially reinstate the relevance of the non-orthogonality condition to the key bits. This is discussed below.

\section{Counterfactual security \label{sec:secu}}

Consider the quantum flip operation $\phi$, which flips the polarization of the photon $H \leftrightarrow V$. In protocol $\mathcal{P}$, the step (2) is replaced by: \\ ($2^\prime$) Alice and Bob independently perform the flip operation on a fraction $f$ instances. In Alice's case, she applies the flip operation to the reflected particle. In Bob's case, he performs the flip operation on the reflected polarization. For example, if he applies $R_H$ followed by a flip, then his action is to block the $V$ polarization and reflect the $H$ polarization, which is then flipped to $V$. We shall refer to this modified protocol as $\mathcal{P}^\prime$. This is also an orthogonal-state based protocol like Noh09, since the flip operation only toggles between two orthogonal states.

By this simple action, they can circumvent Eve's noiseless attack. This is because Eve cannot know when they both did or did not apply the flip operation, each of which would require a different unattack strategy to remove her footprint. To see this, consider a photon of polarization $j$ sent by Alice. The photon's state after Eve's onward attack $\mathcal{U}$ is
\begin{equation}
\ket{\Psi}_{abe} = \frac{1}{\sqrt{2}}(\ket{j,0,\varepsilon_0} + \ket{0,j, \varepsilon_j})_{abe}.
\label{eq:noflip}
\end{equation}
If Alice were to flip and Bob were to apply $R_j$ and introduce the flip as well, then the state of the particle becomes
\begin{equation}
\ket{\Psi^\prime}_{abe} = \frac{1}{\sqrt{2}}(\ket{\overline{j},0,\varepsilon_0} + \ket{0,\overline{j}, \varepsilon_j})_{abe}.
\label{eq:flip}
\end{equation}
For Eve to perfectly disentangle her probe, we should have $\ket{\overline{j}, \varepsilon_j}_{be} \stackrel{\mathcal{U}^{-1}}{\rightarrow}\ket{j, \varepsilon_0}_{be}$. On the other hand, if neither of them performs the flip operation, then we should have $\ket{j, \varepsilon_{j}}_{be} \stackrel{\mathcal{U}^{-1}}{\rightarrow} \ket{j, \varepsilon_0}_{be}$. Clearly, this violates the unitarity of $\mathcal{U}$. Thus, a perfect noiseless attack is impossible.

More general forms of (noisy) attacks by using other probe interaction may be considered, but it is clear that any attempt to correlate Eve's probe with the half-photon $b$ will entangle them, leading to a decoherence observed in the interferometer.

In protocol $\mathcal{P}^\prime$, the definition of QBER in Eq. (\ref{eq:noherror}) should be expanded to:
\begin{subequations}
	\begin{align}
		e_{(1)} &\equiv P(H,R_H|D_1) + P(V,R_V|D_1), \label{eq:errora} \\
		e_{(2)} &\equiv P(H,R_H|D_1, \varphi_{AB}) + P(V,R_V|D_1,\varphi_{AB}), \label{eq:errorb}
	\end{align}
	\label{eq:error}
\end{subequations}
where $e_{(1)}$ (resp., $e_{(2)}$) represents the cases when neither (resp., both) had flipped their reflected polarization in a given $D_1$ detection. $P(\cdot)$ denotes probability and $\varphi_{AB}$ indicates that both Alice and Bob flipped the polarization of the reflected particle. The QBER in the channel is taken to be $e = \max\{e_{(1)}, e_{(2)}\}$.

Thus the step (6) in protocol $\mathcal{P}^\prime$ is replaced by: \\
($6^\prime$) Alice and Bob announce their respective actions that led to a counterfactual detection, to estimate QBER $e$, as given in Eq. (\ref{eq:error}). If $e$ is too large, they abort the protocol run. 

A novel element of our protocol is that the security check is based on a different data (counterfactual bits) than that used for key generation (non-counterfactual bits). This fact gives rise to the curious situation that the key bit is generated at a certain specific place ($D_B$, in Bob’s station), whereas the security checking is accomplished by detections elsewhere ($D_1$, in Alice’s station) produced by IFM. Thus, it seems intuitive to refer to this protection of non-counterfactual bits by counterfactual statistics as \textit{counterfactual security}.

As a simple demonstration of the performance of the protocol, consider Eve's attack where she individually entangles a probe with fraction $g$ of Bob's particles during the onward leg using the unitary $\mathcal{U}$ defined in Eq. (\ref{eq:onward}). To try to make amends for the flip, on the return leg, Eve may perform $\mathcal{U}^{-1}$ or its flipped version $\mathcal{U}_2$ with equal probability. Specifically, $\mathcal{U}_2$ is given by:
\begin{subequations}
	\begin{align}
		\ket{0, x}_{be} &\longrightarrow \ket{0,x}_{be} ~~~(x \in \{\varepsilon_0, \varepsilon_{j}\}), \\
		\ket{j,\varepsilon_j}_{be} &\longrightarrow \ket{j,\varepsilon_j}_{be}, \\
		\ket{\overline{j},\varepsilon_j}_{be} & \longrightarrow \ket{\overline{j},\varepsilon_0}_{be} . 
	\end{align}
	\label{eq:secondg}
\end{subequations} 
When Eve employs $\mathcal{U}^{-1}$, she generates no error if neither Alice nor Bob flip their respective particle, whereas when she employs $\mathcal{U}_2$, she generates no error if both Alice and Bob flip their particle. In either case, an error is generated otherwise, given by $e_{(1)} = e_{(2)} = \frac{g}{(1+g)} \equiv e$. Bob's information is given by $I(A{:}B) = 1 - h(e)$. 

Under the attack, Eve finds the probe in the state $\ket{\varepsilon_{j}}$ and thus acquires deterministic information on all attacked particles registered at $D_B$ (true key bits). Security comes from the fact that when Alice does not announce a $D_1$ detection and Eve finds the probe in this state, Eve cannot decide which of these correspond to false key bits ($D_2$ detections) and which to true key bits ($D_B$ detections). The latter happens on $w \equiv \frac{1}{1 + f(1-f)}$ fraction of all $m \equiv \frac{gn}{4w}$ potential key bits for Eve (wherein she finds the probe in the state $\ket{\varepsilon_{j}}$). Thus, Eve's problem is that she does not know which $\frac{gn}{4}$ of the $m$ potential key bits constitute the true key bits in the fraction of particles she attacked. As there are $\beta \equiv {m \choose mw}$ equi-probable possibilities for the true key string, her ignorance can be quantified by $\log_2(\beta)$. For example, in the case when $f = \half$, $\log_2(\beta) \approx 0.72 \times \frac{5gn}{16}$. On the remaining $(1-g)\frac{n}{4}$ key bits, which she did not attack, Eve is maximally ignorant, so that her ignorance on all key bits is $\frac{n}{4}(0.72 \times \frac{5g}{4} + 1-g) = \frac{n}{4}(1-0.1g)$. Thus, on average $I(A{:}E) = 0.1g$ per key bit.

\begin{figure}[h]
\includegraphics[scale=0.6]{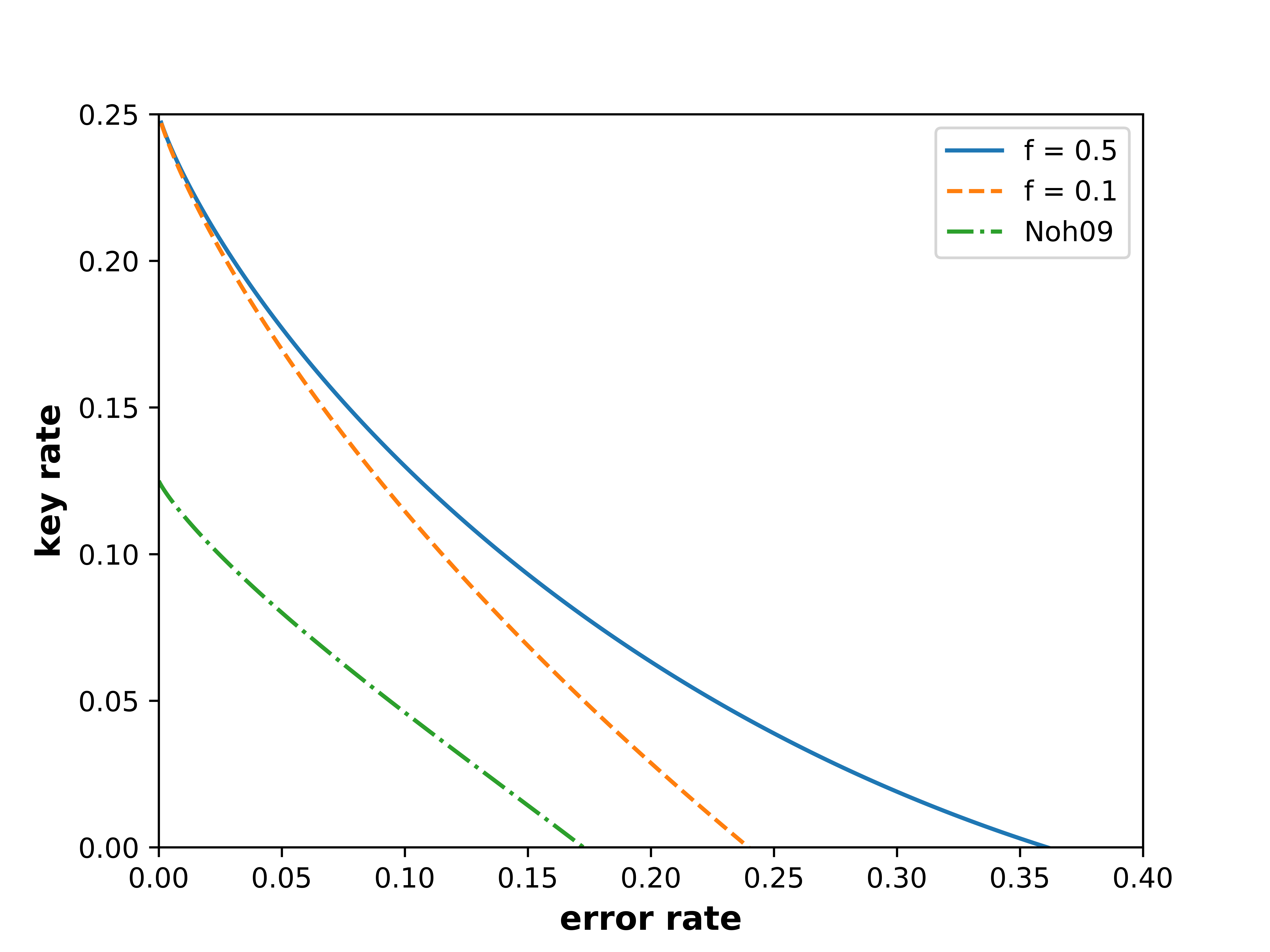}
\caption{(Color online) Key rates as a function of channel error for the proposed protocol and Noh09, both assumed to be subject to a similar incoherent (single-particle) attack. The case of Noh09 is discussed in Section \ref{sec:conc}. The solid and dashed curves pertain to the present protocol, with $f=\half$ and $f=0.1$, respectively. The dot-dashed curve pertains to Noh09, which is found to yield a lower key rate for all $e$ and thus have a lower error tolerance. For the present protocol $f {:=}\half$ is the optimal choice for Alice and Bob, as it maximizes Eve's uncertainty of the true key bits over the potential key bits.}
\label{fig:keyrate}
\end{figure}

The key rate, estimated as $I(A{:}B) - I(A{:}E)$, is then:
\begin{equation}
\kappa(e)\equiv \frac{1}{4} \bigg[1-h(e)-\frac{0.1 e}{1-e}\bigg],
\label{eq:keyrate1}
\end{equation}
and is plotted in Figure \ref{fig:keyrate} for $f=\half$ (solid curve) and $f=0.1$ (dashed curve). For comparison, the key rate of Noh09 under an analogous attack is also given (dot-dashed curve) in the same Figure, showing that the proposed protocol outperforms Noh09 both in terms of key rate and the maximum tolerable error, at least for the considered kind of attack.

In practical terms, the eavesdropper's possible attack can be monitored by the visibility of the interferometer, given by $\gamma \equiv \frac{P(D_2|H,R_H)-P(D_1|H,R_H)}{P(D_2|H,R_H)+P(D_1|H,R_H)}$. Under the above attack, we find 
\begin{equation}
\gamma = 1 - \frac{e}{2(1-e)}
\end{equation}
for the case $f=\half$. This yields $\gamma_{\rm min}= 0.718$ as the minimum tolerable value, corresponding to the error level where the solid curve in Figure \ref{fig:keyrate} drops to zero. 

In Eq. (\ref{eq:keyrate1}), the attack and hence errors generated are symmetric with respect to the cases of both Alice and Bob applying the flip operation as well as neither doing so. In general, the noise generated can evince asymmetry, whereby $e_1 \ne e_2$. A simple instance of this would be when we allow imperfections in the flip operation. Suppose that this operation by Alice and by Bob has a failure probability of $p$. Accordingly, the application of the operation by both parties leads to a $D_2$ detection with probability $(1-p)^2 + p^2$ (when both succeed or both fail), whereas with probability $2p(1-p)$, the flip operation of only one succeeds, and thereby may give rise to a $D_1$ detection (and thus be reflected as an error). Hence, we find that whereas the error $e_{(1)} = 0$, on the other hand $e_{(2)} = \frac{p(1-p)}{1+2p(1-p)} > 0$. 

\section{Other counterfactual QKD protocols \label{sec:other}}

In the semi-counterfactual QKD (SC-QKD) \cite{shenoy2013semi} protocol, Alice injects only photons of a fixed polarization and the encoding is not polarization based. Alice and Bob both have the option to reflect/block the particle. A counterfactual detection will happen only if one of them blocks and the other reflects. Thus the encoding is done by either $\{R_A,B_B\}$ (Alice reflects and Bob blocks) or $\{B_A,R_B\}$ (vice-versa).

The non-counterfactual bits generated by the detection at $D_A$ (Alice's station, at the end of internal arm) or $D_B$ (Bob's station) are discarded in SC-QKD. As in the case of Noh09, in our modification, these non-counterfactual bits are used for key generation, whilst the counterfactual bits are used only to check for errors. In contrast to the original SC-QKD protocol, here Alice publicly announces a detection only if it happens at $D_1$ or $D_2$. This increases the efficiency from $\frac{1}{8}$ to $\frac{1}{2}$, but renders the protocol vulnerable to the noiseless attack, for similar reasons as with protocol $\mathcal{P}$.

As above, this problem is fixed by allowing Alice and Bob to independently apply the flip action on fraction $f$ of the reflected particles. To understand why this helps, suppose without loss of generality that Alice always prepares and sends photons of polarization $H$. In a given instance, if both apply similar flip actions, then the state of the photon after Alice's and Bob's actions becomes \begin{align}
\ket{\Psi^\prime}_{abe} &\stackrel{R_H,R_H}{\longrightarrow} \left\{
\begin{array}{ll} 
\frac{1}{\sqrt{2}}(\ket{H,0,\varepsilon_0} + \ket{0,H, \varepsilon_H})_{abe} & \\
\frac{1}{\sqrt{2}}(\ket{V,0,\varepsilon_0} + \ket{0,V, \varepsilon_H})_{abe} & (\varphi_{AB}),
\end{array}
\right.
\label{eq:scqkd}
\end{align}
given both performed $R_H$ operation. In Eq. (\ref{eq:scqkd}), the first case indicates the flip action by neither of them and the second case, both of them. As is in the case of Noh09, Eve cannot perfectly disentangle her probe because that would require the unattack applied in the return leg to realize both $\ket{\overline{j}, \varepsilon_j}_{be} \rightarrow \ket{j, \varepsilon_0}_{be}$ (\textit{reflect} and \textit{flip}) and $\ket{j, \varepsilon_{j}}_{be} \rightarrow \ket{j, \varepsilon_0}_{be}$ (only \textit{reflect}).

We note that the flip operations only lower the fraction of (counterfactual) check bits, and don't affect the fraction of non-counterfactual key bits. Thus, the efficiency of modified SC-QKD is 
\begin{equation}
\eta_{\rm sc} = P_{D_A} + P_{D_B} = \frac{1}{2}
\label{eq:etasc}
\end{equation}
as $P_{D_A} = P_{D_B} = \frac{1}{4}$. This gives a four-fold increase of efficiency over SC-QKD and doubles with respect to Noh09.

The Guo-Shi protocol \cite{guo1999quantum} works like the SC-QKD scheme but with a Mach-Zehnder setup. As a result, the two-way channel is replaced by two parallel one-way channels. Therefore, the noiseless attack in this case requires a sequential attack by the probe on the two particles. This, however, can be prohibited by the geometry of the setup and carefully monitoring the arrival times. With this additional assumption, non-counterfactual bits can be used for key generation, just as in SC-QKD, but even without introducing flip actions.

In the cascaded version of Noh09 \cite{sun2010counterfactual}, Alice's apparatus is extended by introducing a cascade of $N$ beam-splitters following her first beam splitter. This results in a small amplitude ($2^{-N/2}$) of the photon reaching Bob, thereby exponentially lowering the probability that he can make a $D_B$ detection by his blocking action. Thus, there is little to be gained by including non-counterfactual bits for key generation. However, Alice's counterfactual detections approach the efficiency of $\half$ for very large $N$. In view of Eq. (\ref{eq:etasc}), this efficiency can be achieved by the proposed, much simpler, modified SC-QKD protocol.

\section{Discussion and conclusion \label{sec:conc}}

Noh09 provided an interesting, probabilistic QKD scheme to distribute a secret key conditioned on counterfactual events. Here, we show that non-counterfactual bits in the protocol can be be used for key generation, thereby enhancing the efficiency, whilst the security is still guaranteed by the counterfactual statistics. This leads to the interesting feature of counterfactual security, which accentuates the counterintuitive nature of counterfactuality. 

So far, we assumed that all counterfactual bits are sacrificed as check bits. In principle, a fraction $P_{D_1}^\ast$ of $D_1$ detections can also be used as key bits, when Alice and Bob can establish a secret bit based on their public announcements. The probabilities of this key rate based on the four possible settings is tabulated in Table \ref{tab:past}. 

\begin{center}
	\begin{table}[h]
		\begin{tabular}{| c | c | c | c |} 
			\hline
			\textbf{Flip action by} & {\boldmath$P_{D_1}$} & {\boldmath$\frac{P_{D_1}^\ast}{P_{D_1}}$} \\ [0.5ex] 
			\hline
			Neither & $\frac{1}{8}(1-f)^2$ & $1$ \\ [0.5ex] 
			\hline
			Alice or Bob & $\frac{3}{8}f(1-f)$ & $\frac{1}{3}$ \\ [0.5ex] 
			\hline
			Both & $\frac{1}{8}f^2$ & $1$ \\ [0.5ex] 
			\hline
		\end{tabular}
	\caption{The probability $P_{D_1}$ of $D_1$ detections, and the fraction $P_{D_1}^\ast$ that establishes a secret key, for the four possible flip settings of Alice and Bob.}
	\label{tab:past}
\end{table}
\end{center}
Summing over the second column in Table (\ref{tab:past}), we find 
$
P_{D_1} = \frac{1}{8} + \frac{1}{2}f(1-f).
$
The increase in $D_1$ detections is due to the fact that the flip operations increase the instances of non-interference. Interestingly, the counterfactual key rate $P_{D_1}^\ast=\frac{1}{8}$, equal to the $D_1$ detection probability of Noh09. Thus we find that $\eta_{\rm total} = P_{D_B} + P_{D_1}^\ast = \frac{3}{8}$.
One point to note is that when only one of them flips, the key generation is probabilistic (Table \ref{tab:past}). Thus, in such instances where one party has flipped the polarization, Alice must also announce whether the polarization detected at $D_1$ is consistent with the prepared polarization in order to establish a shared secret bit. This is to take into account the said non-interference and additional information Bob can have to deduce the polarization of the detected photon. Specifically, Alice must announce a consistency (resp. inconsistency) of the polarization of the detected photon when only Alice (resp. Bob) flips the polarization, along with $D_1$ detection instance.

To compare the proposed protocol with Noh09, let us consider the analogue of the individual attack discussed above, in Section \ref{sec:secu}. Directly applying that attack on Noh09 will not work, essentially because of the different methods of key generation in the two protocols. In particular, for counterfactual detections due to Bob's block action (which generate the key bits in Noh09), Eve always finds her probe in state $\ket{\varepsilon_0}$, which ensures that she learns less than Bob. To address this issue, Eve requires to introduce noise in the onward leg to lower the correlation between Alice's and Bob's variables, and in the return leg she requires to minimize the error generated by that action. The simplest modification of the previous attack that achieves this is that in the onward leg, she applies the joint interaction
		\begin{align}
			\ket{j,r}_{be} &\longrightarrow (\cos(\theta)\ket{j,\varepsilon_j} + \sin(\theta)\ket{\overline{j},\varepsilon_j^\prime})_{be} \label{eq:noh2}
		\end{align}
	where the initial state $\ket{0, \varepsilon_0}_{be}$ remains unchanged. Here $\theta$ determines the strength of attack, and $\ket{\varepsilon_0}, \ket{\varepsilon_j}, \ket{\varepsilon_j^\prime}$ are mutually orthogonal probe states.
In the return leg, she applies the joint interaction
		\begin{align}
			\ket{j,\varepsilon_j}_{be} &\longrightarrow \ket{j,\varepsilon_0}_{be}, \nonumber \\
			\ket{\overline{j},\varepsilon_j^\prime}_{be} & \longrightarrow \ket{j,\varepsilon_j^\prime}_{be},
		\end{align}
leaving $\ket{0, x}_{be}$ ($x \in \{\varepsilon_0, \varepsilon_j, \varepsilon_j^\prime\}$) unaltered. 
The error induced by Eve is readily found to be $e_{(1)} = e_{(2)} = \frac{\sin^2(\theta)}{(1 + 2\sin^2(\theta))} \equiv e$, whilst Eve's information
$I(A{:}E) = 2e$ per attacked qubit. Accordingly, we estimate key rate by $\kappa(e) = \frac{1}{8}[1-h(e)-2e]$, which is plotted (dot-dashed curve) in Figure \ref{fig:keyrate}.

This yields a maximum tolerable error of $e_{\rm max} := 17.1\%$, which is lower than that for the proposed protocol (cf. Figure \ref{fig:keyrate}). The corresponding interferometric visibility function is found to be $\gamma = 1 - \frac{\sin^2(\theta)}{2}$, which takes the minimum tolerable value of $\gamma_{\rm min} \approx 0.87$, corresponding to $e_{\rm max}$.

Finally, it may be mentioned that the proposed protocols, including the modified versions discussed in Section \ref{sec:other}, have certain advantages over BB84. For one, the counterfactual QKD protocols are orthogonal state-based, and hence admit simpler state preparations than BB84, which involves non-orthogonal states. Another positive feature is that, unlike in BB84, the check bits are not drawn from the sifted bits, but instead from the data that is eliminated during the sifting process. Finally, we note that the hybrid modified version of SC-QKD, which includes also the counterfactual bits in the key, possesses a higher efficiency (of $\frac{5}{8}$) than BB84 ($\half$).

V.N.R. acknowledges the support and encouragement from Admar Mutt Education Foundation. R.S. is thankful for the support from Interdisciplinary Cyber Physical Systems (ICPS) programme of the Department of Science and Technology (DST), India, Grant No. DST/ICPS/QuST/Theme-1/2019/14.

\bibliography{References}

\end{document}